\documentclass[prl,amsmath,twocolumn,showpacs]{revtex4}
\usepackage{amsmath,graphicx}
\begin{document}
\def\la{{\langle}}
\def\ra{{\rangle}}
\def\a{{\alpha}}
\def\p{\tilde{p}}
\def\x{\tilde{x}}
\def\w{\tilde{W}}
\def\t{\tilde{t}}
\def\k{\tilde{k}}
\def\s{\tilde{\sigma}}

\title{On causality, apparent 'superluminality' and reshaping in barrier penetration}
%
%
\author {D.Sokolovski$^{1,2,3}$}
\address{$^1$ Department of Chemical Physics,  University of the Basque Country, Leioa, Spain\\
$^2$ IKERBASQUE, Basque Foundation for Science\\
$^3$ School of Maths and Physics, Queen's University of Belfast, Belfast, BT7 1NN, UK}

   \date{\today}
   \begin{abstract}
   We consider tunnelling of a non-relativistic particle across a potential barrier.
   It is shown that the barrier acts as 
   an effective beam splitter which builds up the transmitted pulse 
    from the copies of the initial envelope
   shifted in the coordinate space backwards relative to the free propagation.
   Although along each pathway causality is explicitly obeyed, in special cases
   reshaping can result an overall reduction of the initial envelope,
   accompanied by an arbitrary coordinate shift.
   In the case of a high barrier the delay amplitude distribution (DAD) mimics a Dirac 
   $\delta$-function, the transmission amplitude is superoscillatory for finite momenta
   and tunnelling leads to an accurate advancement of the (reduced) initial envelope
   by the barrier width. In the case of a wide barrier, initial envelope is accurately
   translated into the complex coordinate plane. The complex shift, given by the first 
   moment of the DAD, accounts for both the displacement of the maximum of the transmitted
   probability density and the increase in its velocity. It is argued that
   analysing apparent 'superluminality' in terms 
   of spacial displacements helps avoid contradiction associated with time parameters
   such as the phase time.
\end{abstract}

%
%
\pacs{PACS number(s): 03.65.Ta, 73.40.Gk}
\maketitle
\section{I. Introduction}
In 1932 MacColl was first to notice that a wavepacket representing 
a tunnelling particle may emerge from the barrier 
in a manner that suggests that 'there is no appreciable delay in the transmission of the packet through the barrier'  \cite{MCOLL}.
The implication that the particle may have crossed the barrier region with a speed 
greater than the speed of light $c$,
has given the effect the name of 'apparent superluminality'.
A parameter commonly used to estimate the time such a particle spends in the barrier region is
the phase time $\tau_{phase}$, essentially the energy derivative of the phase of the 
transmission amplitude (see, for example \cite{REVS}, \cite{MUGA}).
In accordance with the above, $\tau_{phase}$ becomes independent of the barrier
width $d$ as $d\rightarrow\infty$, a fact often referred to as the Hartman effect \cite{HART}.
Besides tunnelling, a similar behaviour was predicted and observed for a variety of systems, including propagation of a photon through a slab of  'fast light' material, where it has an even more surprising 
aspect, since a free photon already moves at the maximal possible speed $c$ (for a recent review 
 see \cite{Boyd}).
Although it has long been agreed that the the causality is not violated since reshaping
\cite{FOOTRH} destroys
causal relationship between the incident and the transmitted peaks,
exact mechanism of reshaping, the role of the causality principle and
the nature of time parameters used to quantify the effect  remain open to further discussion \cite{Chen}.
With this task in mind, we return here to the case of non-relativistic tunnelling across a potential barrier, originally considered in \cite{MCOLL}.
In \cite{SR0} we analysed a particular type of a beam splitter in which the  transmitted pulse, 
reshaped through interference, appeared reduced and shifted in the coordinate variable relative to free propagation. Post-selection of the particle in a particular spin state allowed one to advance or delay the particle, or to make the shift complex valued. 
With the initial shape of the pulse preserved, the delay amplitude distribution (DAD), which
determines the choice between available pathways, mimicked a Dirac $\delta$-function, while the effective
transmission coefficient exhibited supersocillations \cite{SOSC}, \cite{AHSOSC} in the momentum range of interest.
\newline
The purpose of this paper is to demonstrate that a similar mechanism, albeit without the flexibility
of choosing the delay at will, is realised in non-relativistic tunnelling across a potential barrier.
In Sect.II we change from the momentum to the coordinate representation and show that
the causality principle limits the spectrum of delays available to a transmitted particle.
In Sect.III we show that, due to the oscillatory nature of the complex valued DAD, causality 
alone cannot be used to predict the position of the transmitted pulse.
In Sect.V we analyse advancement in tunnelling across a high rectangular barrier.
In Sect.VI we show that in the semiclassical limit of a wide barrier initial envelope
experiences a complex coordinate shift. In Sect.VII we link the imaginary part of the 
shift to the increase in the mean velocity of the transmitted particle.
In Sect.VIII we explore the analogy between tunnelling and the model of Ref.\cite{SR0} in order to describe the reshaping mechanism.
In Sect. IX we introduce a complex delay time similar to the complex traversal time
\cite{SB} and briefly discuss the wisdom of such an introduction.
Section X contains our conclusions.

\section {II. Delays and causality in one-dimensional scattering.}
Consider, in a non-relativistic limit,  a one-dimensional wave packet with a mean momentum $p_0$ incident from the left on a short-range potential $W(x)$.
Its transmitted part is given by (we put to unity $\hbar$  and the particle's mass $\mu$)
\begin{eqnarray}\label{TS0}
\Psi^T(x,t) =\int T(p) A(p-p_0)\exp(ipx-ip^2t/2) dp
\end{eqnarray} 
where $A(p-p_0)$ is the momentum distribution of the initial pulse,
peaked at $p=p_0$, and $T(p)$ is the transmission amplitude.
Consider also the state $\Psi^0(x,t)$ obtained by free ($W=0$) propagation of 
the same initial pulse,
\begin{eqnarray}\label{TS3}
\Psi^0(x,t) =\int A(p-p_0)\exp(ipx-ip^2t/2) dp.
\end{eqnarray}
It is convenient to extract the phase factor associated with $p_0$ thus defining
functions $G^{T}(x,t,p_0)$ and $G^{0}(x,t,p_0)$ as
\begin{eqnarray}\label{TS6}
G^{T,0}(x,t,p_0) = exp(-ip_0x+ip_0^2t/2)\Psi^{T,0}(x,t).
\end{eqnarray}
Note that $G^{0}(x,t,p_0)$ represents the envelope of the freely propagating state (\ref{TS3}), 
whereas for $\Psi^T(x,t)$, whose mean momentum may have been changed in transmission, it is not,
strictly speaking, so.
Following \cite{SMS} we rewrite the integral (\ref{TS0}) as a convolution 
in the co-ordinate space,  thus obtaining for $G^T$ and $G^0$  in Eq.(\ref{TS6}) 
\begin{eqnarray}\label{TS7}
G^T(x,t,p_0) =T(p_0)\int_{-\infty}^{\infty}\eta(x',p_0)G^0(x-x',t,p_0)dx'.
\end{eqnarray}
In Eq.(\ref{TS7}) $\eta(x)$ is the delay amplitude distribution (DAD), related to the Fourier
transform of the transmission amplitude $T(p)$
\begin{eqnarray}\label{TS2a}
\xi(x) = (2\pi)^{-1} \int_{-\infty}^{\infty} T(p)\exp(ipx)dp,
\end{eqnarray}
as
\begin{eqnarray}\label{TS2}
\eta(x,p_0) = [T(p_0)]^{-1}\exp(-ip_0x) \xi(x),
\end{eqnarray}
and normalised to unity
\begin{eqnarray}\label{TS9a}
\int_{-\infty}^0 \eta(x,p_0)dx = 1.
\end{eqnarray}
Equation (\ref{TS2}), which is exact, demonstrates that at any given time $t$ the transmitted pulse 
$G^T(x,t,p_0)$ builds up from freely propagating envelopes shifted in space by $x'$ (delayed for $x'<0$ and advanced for $x'>0$) and weighted by $\eta(x',p_0)$.
The support of the $\eta(x,p_0)$ (i.e., all $x$ for which $\eta(x,p_0)\ne 0$)
forms a continuum spectrum of available delays.
\newline
The causality principle (CP) ensures analyticity of the transmission amplitude
in the complex $p$-plane \cite{CP} and can be used to obtain information about the spectrum.
In particular, for a barrier potential which does not support
bound states and, therefore, has no poles in the upper half of the complex $p$-plane, $\xi(x)$
must vanish  for $x>0$,  and the spectrum contains no positive shifts (negative delays)  \cite{FOOTCP} .
Accordingly, we can write 
\begin{eqnarray}\label{TS9}
\eta(x,p_0)=\delta(x)+\tilde{\eta}(x,p_0), \quad \tilde{\eta}(x,p_0)\equiv 0, \quad for \quad x>0.
\end{eqnarray}
where the singular term [which arises because $T(p)\rightarrow 1$ for $|x|\rightarrow \infty$]  corresponds to free propagation, while the smooth part  $\tilde{\eta}(x,p_0)$, which describes scattering, vanishes as $W \rightarrow 0$.
Conversely, the CP ensures that for a barrier 
the Fourier transform of $T(p)$ contains only plane waves with non-negative frequencies, $x\ge 0$,
\begin{eqnarray}\label{TS5}
T(p)=\int_0^{\infty}\xi(-x) \exp(ipx)dx.
\end{eqnarray}
Finally, for a barrier we can rewrite Eq.(\ref{TS7}) in an equivalent form
\begin{eqnarray}\label{TS7a}
G^T(x,t,p_0) =\int_{x}^{\infty}\eta(x-x',p_0)G^0(x',t,p_0)dx'
\end{eqnarray} 
which best serves to demonstrate that the CP prevents transfer of information from the tail of the incident pulse to the front of the transmitted one.  Namely, should the envelopes of two freely propagating  wavepackets 
coincide for $x>x_0$, $G_1^0(x,t,p_0)=G_2^0(x,t,p_0)$, then $G_1^T(x,t,p_0)$ and $G_2^T(x,t,p_0)$ will also coincide in the same range,  
making it impossible for an observer to distinguish between the two transmitted pulses until their tails arrive at the detector.
\section{III. Counter-intuitive advancements, superoscillations and quasi-Dirac distributions}
Equation (\ref {TS7}), which is our main result so far,
is worth a brief discussion.
While the overall factor $T(p_0)$ represents a reduction in the magnitude of the transmitted pulse,
its shape is determined by the DAD $\eta(x,p_0)$ and results from 
the interference between the 
sub-envelopes $G^0(x-x',t,p_0)$ with different spacial shifts which,  because a free wavepacket  spreads, depend on time.
The causality principle restricts the spectrum of available shifts and
ensures that in the absence of bound states 
decomposition 
(\ref{TS7}) does not contain advanced terms.
This is a quantum analogue of the classical result that a particle is sped up when passing
over a region where $W(x)<0$, e.g., over a potential well,  and is delayed compared to free propagation
whenever $W(x)>0$, e.g., when passing over a potential barrier.
In the classical limit, $\eta(x,p_0)$ becomes highly oscillatory and has a stationary 
region around $x=x_{cl} $, corresponding to the classical displacement of 
a particle crossing $W(x)$ relative to the free one. 
Thus only one shift $x_{cl}$ and one  shape $G^0(x-x_{cl},t,p_0)$ are selected from those
available in Eq.(\ref{TS7}). Since $\eta(x,p_0)$ must vanish for $x>0$, one can only have $x_{cl} \le 0$. In this way  causality ensures that a particle passing over a barrier can only be delayed.
\newline
Yet when $\eta(x,p_0)$  has no real stationary points, interference effects play the dominant
role and  the CP alone cannot predict the final shape or even the location of the transmitted pulse. 
Indeed, should $T(p)$, for whatever reason,  have a simple exponential 
 form, 
  \begin{eqnarray}\label{R1}
T(p) = B \exp(-i\a p), \quad B=const, \quad \a> 0,
\end{eqnarray} 
 equation (\ref{TS2}) would yield
   \begin{eqnarray}\label{R2}
\eta(x,p_0) = \delta(x-\a),
\end{eqnarray}  
  and the transmitted envelope would be a reduced accurate copy of the freely propagating one,
  advanced by the distance $\a$,
     \begin{eqnarray}\label{R3}
G^T(x,t,p_0) =B G^0(x-\a,t,p_0).
\end{eqnarray} 
  Naively, one may conclude that this situation cannot be realised for a barrier potential,
  given that the CP requires, on one hand, that the Fourier spectrum contain no negative frequencies
  similar to that in Eq.(\ref{R1}) and, on the other hand, that $\eta(x,p_0)$ vanish for all $x>0$ in contradiction to (\ref{R2}).
  However, to achieve the advancement in Eq.(\ref{R3}), its is only necessary that Eqs. (\ref{R1}) 
  and (\ref{R2}) be satisfied approximately \cite{SR0}. Thus, $T(p)$ has to mimic $\exp(-i\a p)$ only in a    
  limited region  of $p$ containing all initial momenta. Equivalently, $\eta(x,p_0)$ has to mimic
  $\delta(x-\a)$ only for initial wavepackets sufficiently broad in the coordinate space.
  The former is possible, since it is well known  \cite{SOSC} that a sum of exponentials, whose   
  frequencies lie within a given interval, can locally reproduce a 'superoscillatory' exponential with a 
  frequency outside this interval. For the latter it is sufficient that the DAD $\eta(x,p_0)$ have several of its 
  moments outside its region of 
  support and equal to 
  those of $\delta(x-\a)$ \cite{SR0}, $\int_{-\infty}^0 x^n \eta(x) dx \approx \a^n$, $n=0,1,2..K$.
  This is possible
  since the DAD is an oscillatory distribution, rather than a non-negative probabilistic one \cite{DSW}.
  If so, the kernel  $\eta(x-x')$  termed in \cite{SR0} a quasi-Dirac distribution, would act like a spacial   
  shift by a distance $\a$ on a polynomial of an order $\le K$ or, more generally, one any function whose 
  Taylor series can be truncated after the first $K$ terms.
  Next we look for evidence of such a behaviour in  tunnelling across a rectangular barrier,
\section { IV. Gaussian wavepackets.} Although the results of Sect. I 
apply, in principle, to initial pulses of arbitrary shape,
in the following we will consider Gaussian wavepackets  with positive momenta incident on the barrier from the left.
Such a  wavepacket has a spacial
width $\sigma$, a mean momentum  $p_0 >0$ and is  centred around some $x=0 $ at $t=0$ so that its momentum distribution $A(p-p_0)$ and the freely propagating envelope in Eq.(\ref{TS7}) are given by
\begin{eqnarray}\label{P2}
A(p-p_0)=\sigma^{1/2}/(2\pi)^{3/4}\times \\
\nonumber
\exp [-(p-p_0)^2\sigma^2/4].
\end{eqnarray}
and
\begin{eqnarray}\label{P1}
G^0(x,t,p_0)=[2\sigma^2/\pi\sigma_t^4]^{1/4}\exp[-(x-p_0t )^2/\sigma_t^2]
\end{eqnarray}
where $\sigma_t^2 \equiv (\sigma^2+2it)$ is a complex valued width which takes into 
account the effects of spreading. The coordinate probability density for the wavepacket
in Eq.(\ref{P1}) has a Gaussian shape
\begin{eqnarray}\label{P3}
\rho^0(x,t)\equiv |G^0(x,t,p_0)|^2=\frac{(2/\pi)^{1/2}}{(\sigma^2-4t^2/\sigma^2)^{1/2}}\times\\
\nonumber
\exp\{-2[x-p_0t]^2/(\sigma^2-4t^2/\sigma^2)\}.
\end{eqnarray}
%
%
%
%

\section{ V. Tunnelling across a high rectangular barrier.}
Consider  tunnelling of a Gaussian wavepacket (\ref{P1}) 
 across a rectangular barrier of a width 
$d$ and a height $W$,  
$W(x)=W$ for $x_0\le x \le x_0+ d$, $x_0>0$, and zero otherwise.
 The transmission amplitude independent of the barrier position $x_0$ is given by 
 \begin{equation}\label{T0a}
T(p,W)=\frac{4pk\exp(-ipd)}{(k+p)^2\exp(-ikd) -(k-p)^2\exp(ikd)}, 
\end{equation} 
where $k=(p^2-2W)^{1/2}$.
It is readily seen that $T(p)$ is single valued in the complex $p$-plane and  has no poles in its upper 
half. The DAD $\eta(x,p_0=0)$ shown in Fig.1 is real because of the symmetry 
$T(-p)=T^*(p)$ and vanishes for $x>0$ as required by causality.
\begin{figure}[h]
\includegraphics[width=8.5 cm, angle=-90]{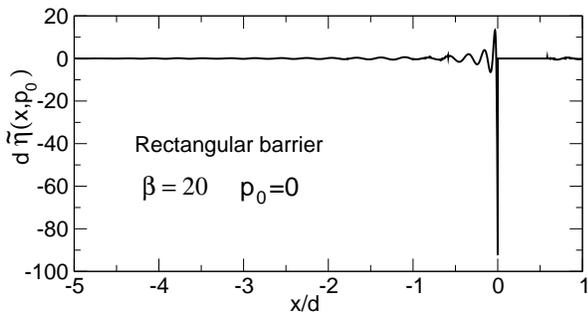}
\vskip0.5cm
\caption{Regular part of the DAD in Eq.(\ref{TS9}) for a rectangular barrier with $\beta \equiv \sqrt{2W}d =
20$ and $p_0=0$ obtained by numerical integration of Eq.(\ref{TS2a}) with $T(p)$ given by Eq.(\ref{T0a}). }
\label{fig:FIG1}
\end{figure}
 It is convenient to
rewrite $T(p)$  as a geometric progression 
\begin{eqnarray}\label{T1}
T(p,W) = \quad \quad \quad \quad \quad \quad \quad  \quad \quad \quad \quad \quad \quad\\
\nonumber
 \frac{4pk \exp[-i(p-k)d]}{(p+k)^2}\sum_{n=0}^{\infty}\frac{(p-k)^{2n}}{(p+k)^{2n}}\exp(-i2nkd),
\end{eqnarray}
where we choose the principal branch of the square root $(p^2-2W)^{1/2}$, i.e.,
$k>0$ for $p^2 > 2W$. 
Next we fix $d$ and the Gaussian momentum distribution of the incident wavepacket $A(p-p_0)$
 and increase the barrier height  so that 
 \begin{eqnarray}\label{T0b}
W \rightarrow \infty, \quad p_0^2/W \rightarrow 0.
 \end{eqnarray}
In this limit it is sufficient to retain  only the $n=0$ term in Eq.(\ref{T1}) and expand it to the leading
 order in $W^{-1}$ to obtain
\begin{eqnarray}\label{T4}
T(p,W)\approx B(W) p \exp(-ip d),
\end{eqnarray} 
with 
\begin{eqnarray}\label{T4a}
B(W) \equiv -4i(2W)^{-1/2}\exp(-\sqrt{2W}d),
\end{eqnarray} 
which can be made valid for all incident momenta.
This is an example of superoscillatory behaviour similar to that discussed in Sect.III. Indeed, $T(p,W)/p$ [regular at $p=0$ according to Eq.(\ref{T0a})], just like $T(p,W)$ itself, has no poles in in the upper half of the complex $p$-plane and its Fourier spectrum cannot contain
negative frequencies. Yet according to Eq.(\ref{T4})  in a limited region around $p=0$ the ratio $T(p)/p$ mimics the behaviour of  $\exp(-ipd)$. 
Further, inserting (\ref{T4}) into Eq.(\ref{TS2}) 
shows that that $\eta(x,p_0)$ mimics the behaviour of a singular distribution with support  at
$x=d$, 
\begin{eqnarray}\label{T5}
\eta(x,p_0) \approx [\delta(x-d)+i\partial_{x}\delta (x-d)/p_0],
\end{eqnarray} 
for a class
 of not-too-narrow wavepackets whose momentum distributions probe only the superoscillatory
 part of $T(p)$.
Accordingly, we find the transmitted pulse reduced in magnitude and advanced 
relative to the free propagation  by the barrier width $d$,
\begin{eqnarray}\label{T2}
G^T(x,t,p_0) \approx T(p_0,W)\times \quad \quad \quad \quad \quad \quad \quad  \quad \quad  \quad \quad \\
\nonumber
[G_0(x-d,t,p_0)-i\partial_xG_0(x-d,t,p_0)/p_0],
\end{eqnarray}
where $T(p_0,W)$ is given by Eq.(\ref{T4}).
There is also an additional distortion term
proportional to $\partial_xG_0(x-d,t,p_0)$, which becomes negligible for sufficiently fast particles. 

Thus, for a given incident Gaussian wavepacket one can always find a barrier high 
 enough, so that the transmitted pulse will be accurately described by Eq.(\ref{T2})
 The price for such an advancement is the reduction of the tunnelling 
 probability by a factor $\sim \exp(-2\sqrt{2W}d)$ which makes transmission a
 very rare event.
 
 The transmission amplitude $T(p)$ and the transmitted pulse $G^T(x,t,p_0)$
 are shown in Fig.2 for the same Gaussian wavepacket and different barrier heights.
 Figure 2 is similar to Fig. 3 of Ref.\cite{SR0} with the difference that for a rectangular
 barrier the superoscillatory band where $T(p)$ can be approximated by Eq.(\ref{T4}) does not 
 have well defined boundaries, whereas for the system studied in \cite{SR0} the transmission 
 amplitude exhibited a much more rapid growth marking the edges of the band.
 Accordingly, the deviations of $G^T(x,t,p_0)$ from the predictions of Eq.(\ref{T2}) at lower
 barrier heights are less pronounced than the distortion of the shape of the transmitted pulse 
 shown in Fig.3b of Ref.\cite{SR0}. 
 \begin{figure}[h]
\includegraphics[width=7 cm, angle=-90]{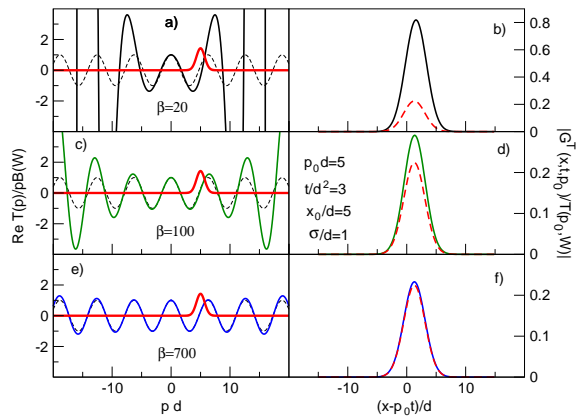}
\vskip0.5cm
\caption{
(Colour online) High rectangular barrier: a) $Re T(p)/p$ (solid) 
and $\sin(-pd)$ (dashed) for
 $\beta \equiv \sqrt{2W}d =20$.
 Also shown is $|A(p-p_0)|$ scaled to a unit height (thick solid).
b) The shape of the transmitted pulse $|G^T(x,p_0,t)/T(p_0,W)|$: exact (solid)
and given by Eq.(\ref{T2}) (dashed); c) and d), same as a) and b) but for $\beta = 100$;
e) and f), same as a) and b) but for $\beta = 700$.
}
\label{fig:FIG2}
\end{figure}
\section {VI.  Tunnelling across a wide rectangular barrier.}
Next we consider the case of tunnelling across a rectangular barrier whose width increases while its height and 
the mean kinetic energy of the particle are kept constant,
\begin{eqnarray}\label{Q0}
d \rightarrow \infty, \quad p_0=const.
\end{eqnarray} 
We will also assume that the width of the incident wavepacket increases proportionally to the barrier width,
\begin{eqnarray}\label{Q1}
\sigma/d\equiv \gamma = const,
\end{eqnarray} 
so that its momentum space width $\sigma_p$ decreases with $d$,
\begin{eqnarray}\label{Q2}
\sigma_p = 2/\sigma = 2/\gamma d.
\end{eqnarray} 
It is easy to show that under these conditions the transmitted pulse
will have the shape of the initial envelope not just advanced
relative to free propagation but also shifted into the complex coordinate plane.
Indeed, retaining only the $n=0$ term in Eq.(\ref{T1}) and expanding the exponent in a Taylor 
series around $p_0$ we may write ($k_0\equiv \sqrt{p_0^2-2W}$)
\begin{eqnarray}\label{Q3}
-id(p-k) = -id\sum_{n=0}^\infty \partial^{n}_p(p-k)|_{p=p_0} (p-p_0)^n/n! \approx  \\
\nonumber
-id(p_0-k_0) -id[1+\frac{ip_0}{\sqrt{2W-p_0}}](p-p_0) 
\end{eqnarray}
for all initial momenta $|p-p_0| \lesssim \sigma_p \sim 1/d$.  We may also replace $p$ with $p_0$ everywhere in 
the pre-exponential factor to finally obtain
\begin{eqnarray}\label{Q4}
 T(p,W) \approx B(p_0,W)  \exp(-ip\a)
\end{eqnarray}
with
\begin{eqnarray}\label{Q5}
\a \equiv d+ip_0d/\sqrt{2W-p_0}
\end{eqnarray}
and 
\begin{eqnarray}\label{Q6}
B(p_0,W) = \frac{4p_0k_0 \exp[-id(p_0-k_0)+i\a p_0]}{(p_0+k_0)^2}.
\end{eqnarray}
Thus, in the range of interest $p_0-\sigma_p \lesssim p  \lesssim p_0+\sigma_p$, $T(p)$ exhibits a kind of a superoscillatory behaviour 
with a complex valued frequency $\a$, $Re \a >0$, similar
to that studied in \cite{SR0}.
As a result, for the transmitted pulse we find
\begin{eqnarray}\label{Q7}
G^T(x,t,p_0) =B(p_0,W) G^0(x-Re \a -i Im \a,t,p_0).
\end{eqnarray} 
With $T(p,W)$ given by Eq.(\ref{Q4}) we expect \cite{SR0}  at least several moments of the DAD
$\eta(x,p_0)$ to equal $\a^n$, $\bar{x^n} \equiv \int x^n \eta(x,p_0) dx =\a^n$, $n=0,1,...$. Using the 
identity
\begin{eqnarray}\label{Q8}
\bar{x^n} =i^n\partial^n_p T(p)/T(p)|_{p=p_0}
, \quad n=0,1,...
\end{eqnarray}
and noting that as $d\rightarrow \infty$ the main contribution to $\bar{x^n}$ comes
from differentiating $n$ times the exponential of the first term in the expansion (\ref{T1}), we obtain
\begin{eqnarray}\label{Q9}
lim_{d\rightarrow \infty} \bar{x^n}/d^n = (\a/d)^n+O(1/d).
\end{eqnarray}
\newline
We can now confirm the result (\ref{Q7}) by repeating the calculation 
in the coordinate space.
Consider, for simplicity, a Gaussian function which can be expanded in a Taylor series
\begin{eqnarray}\label{Q10}
 \exp(-x^2/\gamma^2 d^2)\approx \sum_n  (-1)^n (x/\gamma d)^{2n}/n!.
 \end{eqnarray}
 Inserting (\ref{Q10}) into Eq. (\ref{TS7a}) and using Eq. (\ref{Q9}) we obtain
 ($C^n_k=n!/k!(n-k)!$ is the binomial coefficient)
\begin{eqnarray}\label{Q11}
\int \eta(p_0,x')\exp[-(x-x')^2/\gamma^2 d^2]dx' \approx \quad \quad \quad \\
\nonumber
\sum_n \frac{(-1)^n}{n!(\gamma d)^{2n} }\sum_{k=0}^{2n} C_k^{2n} x^k \bar{x}^{2n-k}\quad \quad \quad 
\\
\nonumber
\approx \exp[-(x-\a)^2/\gamma^2 d^2]+O(1/d) .
\end{eqnarray}
\newline
 Thus, for a given ratio $\sigma/d$ one can always find a barrier wide
 enough, so that the shape of the transmitted pulse will be accurately given by Eq.(\ref{Q7})
 The price for an accurate translation of the freely propagating envelope
 into the complex $x$-plane is the reduction of the tunnelling 
 probability by a factor $|B(p_0,W)|^2$ which makes the transmission a
 very rare event.
 \newline
The ratio between the exact transmission amplitude $T(p)$ 
and the one given by Eq.(\ref{Q4}) as well as
the transmitted envelope $G^T(x,t,p_0)$
 are shown in Fig.3 for the same ratio $\sigma/d$ and different barrier width.
 Figure 3 is similar to Figs.3c and 3f of Ref.\cite{SR0} with the difference that for a rectangular
 barrier the complex superoscillatory band where $T(p)$ can be approximated by Eq.(\ref{Q4}) does not 
 have well defined boundaries. 
 \begin{figure}[h]
\includegraphics[width=7cm, angle=-90]{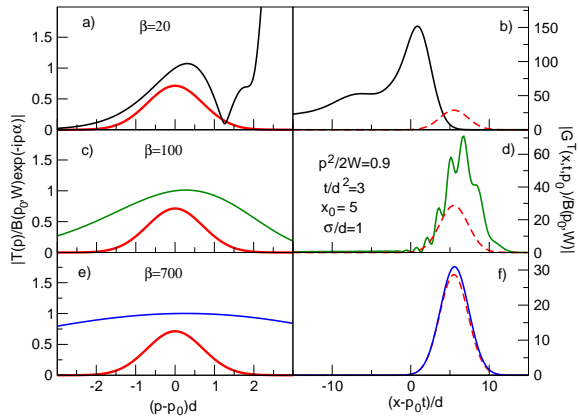}
\vskip0.5cm
\caption{(Colour online) Wide rectangular barrier: a) 
ratio between the exact $T(p)$ and
its approximation in Eq.(\ref{Q4}) for
 $\beta \equiv \sqrt{2W}d =20$ (solid).
 Also shown is $|A(p-p_0)|$ scaled to a unit height (thick solid).
b) The shape of the transmitted pulse $|G^T(x,p_0,t)/T(p_0,W)|$: exact (solid)
and given by Eq.(\ref{Q7}) (dashed); c) and d), same as a) and b) but for $\beta = 100$;
e) and f), same as a) and b) but for $\beta = 700$.}
\label{fig:FIG3}
\end{figure}
\section { VII. Momentum filtering}
Equation (\ref{Q7}) describes, in a compact form, two effects related to the transmission of a 
Gaussian wavepacket. One is a constant shift in the position of the transmitted envelope,
 the other is an increase
 of its average velocity due to suppression of lower momenta contained in the initial distribution.
 Inserting Eq.(\ref{Q4}) and (\ref{P2}) into (\ref{TS0}) and completing the square in the exponent we
have
\begin{eqnarray}\label{U1}
\nonumber
-(p-p_0)^2\sigma^2/4-i\a p =
-(p-p_0-2 Im\a/\sigma^2)^2\sigma^2/4
\\ 
-ipRe \a+p_0 Im\a  +(Im\a)^2/\sigma^2,
\end{eqnarray}
which shows that after the transmission the mean momentum has increased by 
\begin{eqnarray}\label{U1a}
\Delta p_0 = 2p_0d/[\sigma^{2}(2W-p_0^2)^{1/2}].
\end{eqnarray}
Note that $\Delta p_0$ vanishes for a wavepacket very broad in the coordinate
(narrow in the momentum) space. Accordingly, for observable probability density
with the help of  Eqs.(\ref{Q4}) and (\ref{P1}) we find 
\begin{eqnarray}\label{U2}
\rho^T(x,t)\equiv  |G^T(x,t,p_0|^2 =\quad \quad \quad  \quad \quad \quad\quad\quad \quad  \quad \quad \quad\\
\nonumber
C\exp\{-2[x-(p_0+\Delta p_0)t-d]^2/(\sigma^2-4t^2/\sigma^2)\}
\end{eqnarray}
where $C=[2/\pi]^{1/2}\sigma |\sigma_t|^{-2}|B(p_0,W)|^2$
Thus, the transmitted  probability density has a Gaussian shape  which broadens with time and whose maximum propagates
along the trajectory
\begin{eqnarray}\label{U3}
x=(p_0+\Delta p_0)t+d.
\end{eqnarray}
The maximum arrives at a detector earlier than that of a freely propagating pulse [cf. Eq.({\ref{P3})], firstly,
because of the increase in the mean velocity and, secondly, because of additional advancement
by a distance $d$ the pulse has received upon traversing the barrier. This advancement, if 
interpreted incorrectly, gives rise to the notion of 'superluminality'.

\section{VIII. 'Superluminality' and Hartman effect vs. reshaping}
One might try the following classical reasoning: the particle emerges from the barrier with
its mean velocity slightly increased and with an additional advancement. The advancement is due
to a shorter duration $\tau$ spent inside the barrier, $\tau < d/p_0$. Neglecting $\Delta p_0t$, 
for the separation $\Delta x$ between the maxima of the free and the tunnelled pulses one has
$\Delta x = p_0(d/p_0-\tau)$. With $\Delta x =Re \a= d + O(1)$ [c.f. Eq.(\ref{Q9})] we have $\tau \sim 
O(1/p_0)$ and not $O(d/p_0)$ as one might expect. The fact that $\tau$ defined in this manner becomes independent of $d$ in the limit 
of large barrier widths is known as the Hartman effect (see \cite{MUGA} and Refs. therein).
It is readily seen that for a wide barrier $d/\tau$ can be greater than the speed of light $c$,
hence the term 'superluminality' in the title of this Section.
It is well known (see, for example, \cite{MUGA}, \cite{AHSOSC}) that relating the advancement $d$ to 
the duration $\tau$  spent in the barrier is incorrect, since there is no causal relationship between the incident and transmitted peaks. 
\newline
With the help of Eq.(\ref{TS7}) we can analyse reshaping mechanism responsible for destroying
this realtionship.
A barrier 
acts as a beam splitter with an infinite (continuum)
number of arms. On exit from each arm there is an initial pulse shifted backwards
(delayed) by a distance $x'$ and the probability amplitude for passing through the arm 
is $\eta(x',p_0)$. The shifted shapes are then recombined to produce the tunnelled
pulse which, although in none of the arms causality is violated, has an apparently 
'superluminal' aspect.  Resulting wavepacket  is invariably deformed, 
yet it is possible to limit deformation to overall reduction accompanied by 
a coordinate shift, which is 
what happens in the to cases considered in Sects. V and VI. 
Standard quantum mechanics states that if two or more 
different shifts contribute to the sum, no definite shift (delay) can be assigned to the 
product of their interference.
Accordingly, the separation between the free and 
the transmitted  maxima is obtained as the first moment $\bar{x}$ of an alternating complex valued DAD
$\eta(x,p_0)$, for which neither $Re \bar{x}$ nor $Im \bar{x}$ are restricted to lie within 
its region of support \cite{DSW}. Averaging with an amplitude rather than a probability
distribution destroys any direct link between the causal spectrum
of delays in the arms of a beam splitter and apparently non-causal advancement of the transmitted peak.

\section{IX. Complex delays and the phase time}
In the case studied in Sect.IV the observable time parameter of interest is the delay with which
the peak of the transmitted probability density arrives at a detector located at some
$x_d$. With the help of Eq.(\ref{U2}) we can express this delay 
in terms of a complex valued coordinate shift $\a \approx \bar{x}$ which initial Gaussian pulse experiences upon traversing the barrier,  so that  there is no need to introduce any additional time parameters.
If, against our own advice, 
we attribute the coordinate shift $\bar{x}$ to the difference between the durations
$\tau$ and $d/p_0$ spent in the barrier in tunnelling and free motion, 
for $\tau$ we obtain
\begin{eqnarray}\label{W1}
\tau = (d-\bar{x})/p_0 =d/p_0 - i\partial_p \ln T(p_0)/p_0.
\end{eqnarray} Equation (\ref{W1}) defines a complex time parameter, 
whose real part is the phase time \cite{REVS},\cite{MUGA}  often used to quantify advancement of the transmitted pulse,
\begin{eqnarray}\label{W2}
\tau_{phase} \equiv d/p_0+\partial_p \Phi(p_0)/p_0 = Re \tau.
\end{eqnarray}
We note, however, that little is gained by introducing the time parameters (\ref{W1}) and
(\ref{W2}) as  'superluminal' tunnelling is readily analysed in terms of spacial shifts.
It can also be shown that the envelope
plays the role of a pointer in a highly inaccurate 
(weak) quantum measurement of such a shift (see \cite{SMS} and Refs. therein). Both $\tau$ and $\tau_{phase}$ are artefacts of a naive extrapolation of particle-like behaviour to a wave-like situation where, just like in \cite{FOOTRH}, the initial peak is first destroyed and then recreated in a different place by an explicitly causal reshaping mechanism.




%
\section{X. Conclusions and discussion}
In summary, transformation
 to the coordinate representation in Eq.(\ref{TS7}) helps one analyse the reshaping mechanism of quantum tunnelling
as well as the role played by the causality principle.
Like any system characterised by a transmission amplitude $T(p)$, a potential barrier can be seen  as an effective beam splitter with a continuum  of arms (pathways).
On exit from each arm there is a copy of the initial envelope (sub-envelope) shifted relative to free propagation. 
All sub-envelopes recombine to shape the transmitted wavepacket.
The probability amplitude for travelling along a particular pathway is given by the delay amplitude
distribution (\ref{TS2}), essentially a non-analityc Fourier transform of $T(p)$ with an additional 
phase determined by the particle's mean momentum. 
Causality principle
ensures that along neither pathway causality is violated. 
 Thus, for a barrier, non of the
sub-envelopes are advanced, and the Fourier spectrum of a barrier transmission amplitude contains only non-negative frequencies. 
\newline
Restrictions imposed by the CP cannot, however, prevent the reduced tunnelled pulse to be advanced even though all its constituent parts are delayed relative to free propagation.
For example, an accurate advancement by a distance $\a$ is achieved if a sufficient number of 
the complex oscillatory distribution $\eta(x,p_0)$ equal $\a^n$, $n=0,1,..$ where $\a$ lies outside
the spectrum of available shifts. Equivalently, in a limited region of momenta,  $T(p)$ mimics  the exponential $\exp(-i\a p)$ with a frequency outside its Fourier spectrum.
The width of this superoscillatory band imposes the limit on the minimal coordinate width of 
a wavepacket  which can be advanced without distorting the shape of the envelope.
\newline 
For a rectangular barrier of the width $d$, one can find at least two regimes where 
a situation similar to the one just described is realised. 
Well above the barrier a single shift is selected from the spectrum 
and one can speak, in a classical sense, of a duration spent in the barrier region.
Whenever more than two sub-envlopes envelopes interfere, no such duration can be assigned
to the distorted (reshaped) transmitted pulse. In the case of a high barrier considered in Sect.V , this distortion takes the
form of an overall reduction in size accompanied by forward shift by the barrier width $d$.
In the special case of tunnelling across a wide barrier considered in Sect.VI the distortion takes the form of an overall reduction accompanied by a complex valued coordinate shift $\a$.  The shift accounts for the shift of the maximum of the transmitted probability density as well as for the increase in its velocity. Since the free Hamiltonian commutes with a coordinate shift, whether real or complex, the above remains true at any time, once the transmission is completed.
This analysis can be compared with the description of the effect in terms of the phase time (\ref{W2}):
were the transmitted pulse (\ref{Q7}) to represent, ( which it doesn't), a classical particle crossing the barrier region, such a particle would have to cross a wide barrier infinitely fast. Arguably, the latter statement raises more questions then provides answers and contributes to the extended discussion of the subject which continues in the literature \cite{REVS} .
\newline
The origin of $\a$ is of some interest. The shift $\a \approx \bar{x}$  is the complex-valued first moment of the alternating delay amplitude distribution $\eta(x,p_0)$. It has been shown in Ref. \cite{DSW} that such non-probabilistic averages arise whenever one attempts to answer the 'which way?'
(in our case, 'which shift?') question without destroying interference between different
pathways. Standard quantum mechanics cannot give (and, according to \cite{Feyn2} 
best avoid trying to give)  a consistent answer to this question,
and the over-interpretation of the 'weak value' $\a$ leads to a false notion of 'superluminarity' as discussed above.
\newline
Finally, our analysis  applies to a wavepacket of an arbitrary shape with
a sufficiently  narrow momentum distribution.  The Gaussian wavepackets considered above have an additional advantage of being sufficiently well localised in both the coordinate and the momentum spaces and, for this reason, provide a good illustration of the quantum speed up effect.



\begin{thebibliography}{999}
\bibitem{FOOTRH} A trivial yet instructive example of reshaping consists in preparing a piece of paper in a shape of, say, equilateral triangle, using scissors to cut out a smaller triangle of the same shape from its front part and discarding the rest. The peak of the triangle would undergo an instantaneous advancement, which, needless to say, is in perfect agreement with relativistic causality.
\bibitem{FOOTCP} The role of the causality principle becomes clearer in the case of a 
light pulse propagating, say, across a slab of a fast-light material \cite{Boyd}.
In the absence of bound state in which a photon can be captured, the sub-envelopes that build up
the transmitted pulse cannot be advanced relative to the free propagation at 
the maximum possible speed $c$.
\end{thebibliography}
\end{document}